\documentclass[11pt]{article}
\usepackage[latin9]{inputenc}
\usepackage{amsfonts}
\usepackage{amsmath}
\usepackage{amssymb}
\usepackage{indentfirst}
\usepackage{graphicx}
\usepackage[colorlinks]{hyperref}
\usepackage{cite}
\usepackage{graphicx}

\numberwithin{equation}{section}
\begin{document}

\begin{titlepage}
\vspace{3cm}
\baselineskip=24pt

\begin{center}
\textbf{\LARGE{Generalized cosmological term in $D=4$ supergravity from a new $AdS$-Lorentz superalgebra}}
\par\end{center}{\LARGE \par}

\begin{center}
	\vspace{1cm}
	\textbf{Diego M. Pe\~{n}afiel}$^{\ast}$,
	\textbf{Lucrezia Ravera}$^{\ddag}$
	\small
	\\[5mm]
	$^{\ast}$\textit{Instituto
		de Física, Pontificia Universidad Católica de Valparaíso, }\\
	\textit{ Casilla 4059, Valparaiso-Chile.}
	\\[2mm]
                 $^{\ddag}$\textit{INFN, Sezione di Milano, }\\
	\textit{ Via Celoria 16, I-20133 Milano-Italy.}
	\\[5mm]
	\footnotesize
	\texttt{diego.molina.p@pucv.cl},
	\texttt{lucrezia.ravera@mi.infn.it}
	\par\end{center}
\vskip 20pt
\begin{abstract}
\noindent

A new supersymmetrization of the so-called $AdS$-Lorentz algebra is presented. It involves two fermionic generators and is obtained by performing an abelian semigroup expansion of the superalgebra $\mathfrak{osp}(4|1)$. 
The peculiar properties of the aforesaid expansion method are then exploited to construct a $D=4$ supergravity action involving a generalized supersymmetric cosmological term in a geometric way, only from the curvatures of the novel superalgebra. The action obtained with this procedure is a MacDowell-Mansouri like action. Gauge invariance and supersymmetry of the action are also analyzed.

\end{abstract}
\end{titlepage}\newpage {}

\section{Introduction}


As it is well known, a good candidate for describing dark energy is the cosmological constant (see, for example, \cite{Frieman:2008sn, Padmanabhan:2008if}). Then, it becomes interesting to analyze the ways in which cosmological constant terms can be introduced in (super)gravity theories. In particular, the $D=4$ supergravity theory with a cosmological term can be developed in a geometric formulation, where the theory is construted from the curvatures of the superalgebra $\mathfrak{osp}(4|1)$ and the resulting action is the so-called MacDowell-Mansouri action \cite{MacDowell:1977jt}. 

{On the other hand, as it was shown in \cite{Miskovic:2009bm, Miskovic:2014zja}, the renormalized action for $AdS$ gravity in four-dimensions, corresponding to the (bosonic) MacDowell-Mansouri action, is on-shell equivalent to the square of the Weyl tensor describing
conformal gravity. 
Then, a further motivation in the construction of (super)gravity MacDowell-Mansouri like actions lies in the fact that they suggest a (super)conformal structure.}\footnote{{Let us mention, here, that the development of more complicated supergravity theories, such as matter-couple ones, can be carried on in a systematic way with the superconformal method (see the book \cite{Freedman:2012zz} for a review). Indeed, in this approach, which is based on early work \cite{Kaku1, Kaku2, Kaku3, Kaku4} on the understanding of supergravity from the superconformal algebra \cite{Haag:1974qh}, the conformal symmetry is properly used as a tool for constructing matter-coupled supergravity theories with local Poincar\'{e} supersymmetry in such a way to gain insight into their structure.}}


Considering, instead, Minkowski spacetime, its symmetries are described by the Poincar\'{e} algebra, which is generated by $\lbrace J_{ab}, P_a \rbrace$, being $J_{ab}$ and $P_a$ the Lorentz and spacetime translations generators, respectively{, where, in particular,}
\begin{equation}
{[P_a, P_b] = 0.}
\end{equation}
The symmetries of Minkowski spacetime can be generalized and extended from the Poincar\'{e} to the Maxwell symmetries {\cite{Bacry:1970ye, Schrader:1972zd, Beckers:1983gp, Soroka:2004fj, Bonanos:2008kr, Bonanos:2008ez, Gomis:2009dm, Gomis:2009vm, Bonanos:2010fw, Gibbons:2009me}} (see also the more recent work \cite{Concha:2016hbt} for an alternative way of closing Maxwell-like algebras). 
{The most general deformation is
\begin{equation}\label{commutator1}
[P_a, P_b] = Z_{ab} , 
\end{equation}
where $Z_{ab}$ ($Z_{ba} = - Z_{ab}$) transforms as a tensor under Lorentz transformations and is associated with a constant electromagnetic background field.
Then, one can analyze the commutator of $Z_{ab}$ with $P_a$ and iterate the procedure \cite{Bonanos:2008ez}. In this way, as it was shown in \cite{Gomis:2017cmt} (see also \cite{Gomis:2018xmo}), the most general structure one obtains is the infinite-dimensional \textit{free Lie algebra}, called Maxwell$_\infty$ in the quoted references, generated by the $P_a$'s.}

{The (``conventional'') Maxwell algebra $\mathcal{M}$ of \cite{Schrader:1972zd, Beckers:1983gp, Soroka:2004fj, Bonanos:2008ez} (also denoted in the literature as $\mathfrak{B}_4$)} is generated by the set $\lbrace J_{ab}, P_a, Z_{ab} \rbrace$ and its commutation relations read as follows:
\begin{equation}\label{maxwell}
\begin{split}
\left[ P_a, P_b \right] & = \Lambda Z_{ab}, \\
\left[ J_{ab},J_{cd}\right] & =\eta _{bc}J_{ad}-\eta _{ac}J_{bd}-\eta
_{bd}J_{ac}+\eta _{ad}J_{bc}, \\
\left[ J_{ab},P_{c}\right] & = \eta_{bc} P_a -\eta_{ac} P_b , \\
\left[ J_{ab},Z_{cd}\right] & =\eta _{bc}Z_{ad}-\eta _{ac}Z_{bd}-\eta
_{bd}Z_{ac}+\eta _{ad}Z_{bc}, \\
\left[ Z_{ab},Z_{cd}\right] & = 0 , \quad \left[ Z_{ab},P_{c}\right]  = 0 .
\end{split}
\end{equation}
{Here, the bosonic generators $Z_{ab}$ are tensorial abelian charges. We stress that the Maxwell algebra is not unique at bosonic (and supersymmetric) level. From the Lie algebra cohomology perspective, one can deform also the commutation relations $\left[ Z_{ab},Z_{cd}\right] =0$ and $\left[ Z_{ab},P_{c}\right]  = 0$ requiring the Jacobi identity to be obeyed in order to end up with a consistent Lie algebra \cite{Bonanos:2008ez, Gomis:2017cmt, Gomis:2018xmo}, and iterate the construction to properly obtain all possible multiple commutators of the $P_a$'s (namely, the so-called \textit{free Lie algebra} generated by the $P_a$'s).}

The constant $\Lambda$ {appearing in \eqref{maxwell}} can be related to the cosmological constant when $[\Lambda ] = M^2$. Setting $\Lambda = e$, being $e$ the electromagnetic coupling constant, we get the description of an enlarged spacetime in the presence of a constant electromagnetic background field. Indeed, in order to interpret the Maxwell algebra (and the corresponding Maxwell group), a Maxwell group-invariant particle model on the extended spacetime $(x^\mu, \phi^{\mu \nu})$, with the translations of $\phi^{\mu \nu}$ generated by $Z_{\mu \nu}$ (that is $Z_{ab}$ in the components language), was studied {\cite{Bonanos:2008kr, Bonanos:2008ez, Gomis:2009dm, Gomis:2009vm, Bonanos:2010fw, Gibbons:2009me}}: The interaction term described by a Maxwell-invariant $1$-form introduces new tensor degrees of freedom $f_{\mu \nu}$, momenta conjugate to $\phi^{\mu \nu}$, that, in the equations of motion, play the role of a background electromagnetic field which is constant on-shell and leads to a closed, Maxwell-invariant $2$-form.
Interestingly, in \cite{deAzcarraga:2010sw} the authors presented an alternative way of introducing the generalized cosmological constant term adopting the Maxwell algebra.


On the other hand, it was shown that the Maxwell symmetries can be deformed in such a way to obtain the $\mathfrak{so}(D-1,2) \oplus \mathfrak{so}(D-1,1)$ or $\mathfrak{so}(D,1)\oplus \mathfrak{so}(D-1,1)$ algebras {\cite{Gomis:2009dm, Bonanos:2010fw, Soroka:2006aj, Durka:2011nf}}, where the $Z_{ab}$ generators are non-abelian. Then, if spacetime symmetries are considered as local symmetries, one can construct Chern-Simons gravity actions in which dark energy can be interpreted as part of the metric of spacetime. 

Subsequently, in \cite{Salgado:2014qqa} it was shown that a generalized cosmological constant term can also be introduced in a Born-Infeld like action constructed from the curvatures of the so-called $AdS$-Lorentz algebra, $AdS-\mathcal{L}_4$ (also known as $\mathfrak{so}(D-1,1) \oplus \mathfrak{so}(D-1,2)$, or Poincar\'{e} semisimple extended algebra). 
{The $AdS$-Lorentz algebra, which was first introduced and described as a tensorial semi-simple enlargement of the Poincar\'{e} algebra and as a Lie algebra deformation of the Maxwell algebra {\cite{Soroka:2004fj, Gomis:2009dm, Bonanos:2010fw, Soroka:2006aj, Durka:2011nf}}, can also be obtained by performing an abelian semigroup expansion ($S$-expansion, for short) of the $AdS$ algebra, as it was shown in \cite{Diaz:2012zza}.}
{The general theory of expansions of Lie (super)algebras, which allows to derive new Lie (super)algebras and, correspondingly, new physical theories, was first introduced in \cite{deAzcarraga:2002xi}. Subsequently, in \cite{Izaurieta:2006zz} the authors formulated the so-called $S$-expansion procedure, which is based on combining the multiplication law of a semigroup $S$ with the structure constants of a Lie (super)algebra $\mathfrak{g}$. The new Lie (super)algebra one ends up with is called the $S$-expanded (super)algebra $\mathfrak{g}_S = S \times \mathfrak{g}$, and the $S$-expansion method also provide in a simple way an invariant tensor for it, in terms of an invariant tensor for the starting (super)algebra $\mathfrak{g}$, which is particularly helpful in the (geometric) construction of new (super)gravity theories.}
In fact, diverse (super)gravity theories have been studied by exploiting the $S$-expansion method and its properties (see, for instance, the relevant results presented in \cite{Izaurieta:2006aj, Izaurieta:2009hz, Concha:2013uhq, Fierro:2014lka, Concha:2014vka, Concha:2014zsa, Concha:2014tca, Concha:2015tla, Concha:2016kdz, Concha:2016tms, Concha:2016zdb, Caroca:2017onr, Caroca:2017izc, Concha:2018zeb, Concha:2018jxx}).

{The $S$-expansion method allows to obtain two separate types of Lie algebras, denoted in the literature as $\mathfrak{B}_m$ (Maxwell algebras type) and $AdS-\mathcal{L}_m$ (for a concise review see, for example, \cite{Concha:2016hbt}). Both can be related with each other by In\"{o}n\"{u}-Wigner
contraction. The integer index $m > 2$ labels different representatives, where standard generators of the Lorentz transformations $J_{ab}$ and translations $P_a$ become equipped with another set (or sets) of new generators $Z_{ab}$ and $R_{a}$. The value $m-1$ might be used to indicate the total number of different generators ($J_{ab}, P_a, Z_{ab} = Z_{ab}^{(1)}, R_a, = R_a^{(1)}, Z_{ab}^{(2)}, R_a^{(2)}, \ldots$). In particular, $\mathfrak{B}_3$ and $AdS-\mathcal{L}_3$ correspond to the Poincar\'{e} and to the $AdS$ algebras, respectively, while $\mathfrak{B}_4$ represents the Maxwell algebra $\mathcal{M}$ written in \eqref{maxwell} (obtained by including the $Z_{ab}$ generator).}


Referring to superalgebras, the minimal supersymmetrization of the Maxwell algebra ($s\mathcal{M}$) was introduced in \cite{Bonanos:2009wy}, and it requires two fermionic charges.
Extensive studies and analyses of the minimal Maxwell superalgebra and its generalizations have been preformed using expansion methods in \cite{deAzcarraga:2012zv, Concha:2014xfa}. 
Let us mention that other relevant superalgebras containing two fermionic charges were introduced and deeply analyzed in \cite{DAuria:1982uck, Green:1989nn, Andrianopoli:2016osu, Andrianopoli:2017itj, Penafiel:2017wfr, Ravera:2018vra}. In particular, we are referring to the D'Auria-Fr\'{e} superalgebra, introduced in \cite{DAuria:1982uck} and further analyzed in \cite{Andrianopoli:2016osu, Andrianopoli:2017itj}, underlying the Free Differential Algebra describing $D=11$ supergravity, the Green algebra \cite{Green:1989nn}, in the context of superstring, and Maxwell-type superalgebras related to $D=4$ and $D=11$ supergravity \cite{Penafiel:2017wfr, Ravera:2018vra}. 

Concerning the minimal Maxwell superalgebra $s\mathcal{M}$, in \cite{deAzcarraga:2014jpa} it was shown that the $\mathcal{N} = 1$, $D = 4$ pure supergravity Lagrangian can be obtained as a quadratic expression in the curvatures associated with $s\mathcal{M}$. The action of \cite{deAzcarraga:2014jpa} does not include the cosmological constant.

{Let us specify, here, that in the recent paper \cite{Gomis:2018xmo} the authors gave the structure of a \textit{free Lie superalgebra} for studying extensions of the Poincar\'{e} superalgebra, and that the contractions of the $\mathfrak{B}_m$ algebras we have considered so far are subalgebras of the free Lie (Maxwell super)algebra, which can be viewed as a universal structure comprising, by quotienting, generalizations existing in the literature. In particular, as shown in \cite{Gomis:2018xmo}, the superalgebra $s\mathcal{M}$ arises as a particular finite-dimensional quotient of the free Lie superalgebra in four spacetime dimensions generated by odd supertranslations $Q_\alpha$.}


On the other hand, in \cite{Concha:2015tla} the authors demonstrated that the $AdS$-Lorentz superalgebra $sAdS-\mathcal{L}_4$ (minimal supersymmetrization of the $AdS$-Lorentz algebra $AdS-\mathcal{L}_4$ of \cite{Salgado:2014qqa}) allows to construct in a geometric way the supergravity containing a generalized supersymmetric cosmological constant.
The four-dimensional action of \cite{Concha:2015tla} is built only from the curvatures of $sAdS-\mathcal{L}_4$ and corresponds to a MacDowell-Mansouri like action. 
In the same paper, the authors also extended their result introducing the so-called generalized minimal $AdS$-Lorentz superalgebra, building a more general action.

The superalgebra $sAdS-\mathcal{L}_4$ of \cite{Concha:2015tla} presents the following anticommutation relation:
\begin{equation}
\left\{ Q_{\alpha }, Q_{\beta }\right\}  = - \frac{1}{2} \left[ \left(\gamma^{ab} C \right)_{\alpha \beta} Z_{ab} - 2 \left( \gamma ^{a}C\right) _{\alpha \beta }P_{a} \right],
\end{equation}
being $Q_\alpha$ a four-components Majorana spinor charge.
In $(s)AdS-\mathcal{L}_4$, unlike the case of the (minimal supersymmetrization of) the Maxwell algebra $(s)\mathcal{M}$, the new generators $Z_{ab}$ are non-abelian and behave as Lorentz generators. Their presence implies the introduction of a new bosonic field which modified the definition of the curvatures when building a (super)gravity theory for the $AdS$-Lorentz (super)algebra $(s)AdS-\mathcal{L}_4$. Furthermore, $sAdS-\mathcal{L}_4$ contains only one spinor charge, while, as shown in \cite{Bonanos:2009wy}, the minimal supersymmetrization of the Maxwell algebra requires two Majorana spinor charges.

Actually, as we have already mentioned, in \cite{Concha:2015tla} the authors also presented the generalized minimal $AdS$-Lorentz superalgebra, which involves two fermionic charges. However, it also includes two extra bosonic generators ($\tilde{Z}_{ab}$ and $\tilde{Z}_a$), and an In\"{o}n\"{u}-Wigner contraction of the generalized minimal $AdS$-Lorentz superalgebra provides the generalized minimal Maxwell superalgebra $s\mathcal{M}_4$ \cite{Concha:2014tca}, which involves an extra bosonic generator $\tilde{Z}_{ab}$ with respect to the minimal super-Maxwell algebra $s\mathcal{M}$ of \cite{Bonanos:2009wy}. Then, in order to end up with the Maxwell algebra $\mathcal{M}$, one should further contract the bosonic subalgebra of $s\mathcal{M}_4$ in such a way to remove the extra bosonic generator $\tilde{Z}_{ab}$.

{The} $AdS$-Lorentz type (super)algebras also possesses the non-commutativity $\left[ P_a , P_b \right] = Z_{ab}$, which is also present in the Maxwell (super)symmetries.
Nevertheless, unlike the MacDowell-Mansouri Lagrangians for $\mathfrak{osp}(4|1)$ and for $sAdS-\mathcal{L}_4$, it was shown in \cite{Concha:2014tca} that the supergravity action \`{a} la MacDowell-Mansouri based on the generalized minimal Maxwell superalgebra $s\mathcal{M}_4$ does not reproduce the supersymmetric cosmological constant term in the action. This is a direct consequence of the $S$-expansion procedure. The result of \cite{Concha:2014tca} corresponds to a generalization of the one previously presented in \cite{deAzcarraga:2014jpa}. {In both cases,} the super-Maxwell fields only appears in boundary terms of the resulting Lagrangian.

{Let us specify that, in general, the (super-)Maxwell (super)gravity theories are not invariant under the corresponding Maxwell (super)algebra. On the other hand, there has been a growing interest in the study of three-dimensional Chern-Simons (super)gravity theories invariant under Maxwell type (super)algebras \cite{Concha:2016zdb, Caroca:2017izc, Concha:2018jxx, Concha:2018zeb, Salgado:2014jka, Hoseinzadeh:2014bla, Aviles:2018jzw, Concha:2015woa}. In particular, in \cite{Concha:2018jxx} the authors presented the construction of the $(2+1)$-dimensional Chern-Simons supergravity theory invariant, by construction, under the minimal Maxwell superalgebra (that is, under the super-Maxwell gauge transformation) in absence of extra fields, obtaining a supergravity action without cosmological
constant term characterized by three coupling constants, and showed that the
Maxwell supergravity obtained appears as a vanishing cosmological constant limit of a
minimal $AdS$-Lorentz supergravity (the flat limit was applied at the level of the superalgebra,
Chern-Simons action, supersymmetry transformation laws, and field equations).}

Thus, introducing a generalized supersymmetric cosmological term through super-Maxwell symmetries in the context of supergravity seems to be a hard task.\footnote{Actually, this was done in \cite{Durka:2011gm}, but in the formalism of the so-called contrained BF theories (see \cite{Durka:2011gm} and references therein).}
Conversely, the $AdS$-Lorentz type superalgebras seem to be better candidates for introducing the cosmological term in supergravity, in the presence of bosonic generators $Z_{ab}$. 
Let us also mention that, in \cite{Concha:2015tla}, unlike the case of Maxwell-type superalgebras, the bosonic fields associated to the Lorentz-like generators $Z_{ab}$ appear not only in the boundary terms but also in the bulk Lagrangian of the model.


With this motivation, the aim of the present paper it to construct a MacDowell-Mansouri like action based on a new $AdS$-Lorentz type superalgebra involving two fermionic charges and possessing a bosonic subalgebra that does not contains any additional generator with respect to the Maxwell algebra $\mathcal{M}$, in such a way that, when contracted, it directly reproduces $\mathcal{M}$.
In this sense, such a superalgebra could also be viewed as the minimal supersymmetrization of a minimal Maxwell-like algebra {(actually, of a deformation of $\mathcal{M}$)} in which the bosonic generator $Z_{ab}$ is non-abelian (in particular, $\left[ Z_{ab},Z_{cd}\right]  = \eta _{bc}Z_{ad}-\eta _{ac}Z_{bd}-\eta _{bd}Z_{ac}+\eta _{ad}Z_{bc}${, see also \cite{Gomis:2009dm, Bonanos:2010fw} for non-abelian deformations of Maxwell (super)algebras)} and where $\left[ Z_{ab},P_{c}\right] = \eta _{bc}P_{a}-\eta _{ac}P_{b}$, even if, as we will see, the generators $Z_{ab}$ in this case will not behave as Lorentz generators when considering the supersymmetric extension and the corresponding commutation relations with the fermionic charges.
Thus, in this paper we first present the aforesaid new supersymmetrization of $AdS-\mathcal{L}_4$ (we will call it $\tilde{s}AdS-\mathcal{L}_4$, for short) as an abelian semigroup expansion of $\mathfrak{osp}(4|1)$. Then, we show that $\tilde{s}AdS-\mathcal{L}_4$ allows to construct in a geometric way a $D=4$ supergravity action containing a generalized supersymmetric cosmological term.
The action we end up with corresponds to a MacDowell-Mansouri like action written in terms of the $\tilde{s}AdS-\mathcal{L}_4$ curvatures. Our result is a new supersymmetric extension of \cite{Salgado:2014qqa} involving two fermionic generators.
In our model, the $\tilde{s}AdS-\mathcal{L}_4$ fields will appear not only in the boundary terms, but also in the bulk Lagrangian (analogously to what happened in \cite{Concha:2015tla} and differently from what happened in \cite{deAzcarraga:2014jpa, Concha:2014tca}). Referring to the recent works \cite{Andrianopoli:2014aqa, Ipinza:2016con, Banaudi:2018zmh, Concha:2018ywv}, we conjecture that the presence of the $\tilde{s}AdS-\mathcal{L}_4$ fields in the bulk and in the boundary would allow to recover the supersymmetry invariance of a supergravity theory based on the superalgebra $\tilde{s}AdS-\mathcal{L}_4$ in the presence of a non-trivial boundary of spacetime in the so-called rheonomic (i.e.{,} geometric) approach.

The paper is organized as follows: In Section \ref{S1}, we present the new superalgebra $\tilde{s}AdS-\mathcal{L}_4$ as an $S$-expansion of $\mathfrak{osp}(4|1)$. Subsequently, in Section \ref{S2} we construct in a geometric way a $D=4$ supergravity model containing a generalized supersymmetric cosmological term only from the curvatures of $\tilde{s}AdS-\mathcal{L}_4$; the action obtained corresponds to a MacDowell-Mansouri like action. Then, we analyze the supersymmetry invariance of the theory. Finally, Section \ref{Discussion} contains the conclusions and possible future developments. In the Appendix we collect our conventions and some useful formulas.

\section{$AdS$-Lorentz superalgebra $\tilde{s}AdS-\mathcal{L}_4$ as an $S$-expansion of $\mathfrak{osp}(4|1)$}\label{S1}

In the following, we apply the $S$-expansion method to $\mathfrak{osp}(4|1)$ by using a particular abelian semigroup and obtain a new supersymmetrization of $AdS-\mathcal{L}_4$. We will name this novel superalgebra $\tilde{s}AdS-\mathcal{L}_4$, for short. 

Let us first recall that the generators $\lbrace \tilde{J}_{ab}, \tilde{P}_a, \tilde{Q}_\alpha \rbrace$ (with $a=0,1,2,3$, $\alpha=1,2,3,4$) of $\mathfrak{osp}(4|1)$ fulfill the (anti)commutation relations:\footnote{Here and in the following we denote the quantities referring to $\mathfrak{osp}(4|1)$ with a tilde symbol on the top.}
\begin{equation}\label{osp}
\begin{split}
\left[ \tilde{J}_{ab}, \tilde{J}_{cd}\right] & =\eta _{bc}\tilde{J}_{ad}-\eta _{ac}\tilde{J}_{bd}-\eta
_{bd}\tilde{J}_{ac}+\eta _{ad}\tilde{J}_{bc}, \\
\left[ \tilde{J}_{ab},\tilde{P}_{c}\right] & =\eta _{bc}\tilde{P}_{a}-\eta _{ac}\tilde{P}_{b}, \\
\left[ \tilde{P}_a , \tilde{P}_b \right]  & =\tilde{J}_{ab}, \\
\left[ \tilde{J}_{ab},\tilde{Q}_{\alpha }\right] & =-\frac{1}{2}\left( \gamma _{ab}\tilde{Q}\right)_{\alpha }, \\
\left[ \tilde{P}_{a}, \tilde{Q}_{\alpha }\right] & =-\frac{1}{2} \left( \gamma _{a}\tilde{Q}\right) _{\alpha }, \\
\left\{ \tilde{Q}_{\alpha }, \tilde{Q}_{\beta }\right\} & = - \frac{1}{2} \left[ \left(\gamma^{ab} C \right)_{\alpha \beta} \tilde{J}_{ab} - 2 \left( \gamma ^{a}C\right) _{\alpha \beta }\tilde{P}_{a} \right] ,
\end{split}
\end{equation}
where $\gamma_{ab}$, $\gamma_a$ are Dirac gamma matrices in four dimensions and $C$ is the charge conjugation matrix; $\tilde{J}_{ab}$ are the Lorentz generators, $\tilde{P}_a$ the spacetime translations generators, and $\tilde{Q}_\alpha$ is a four-components Majorana spinor charge.

Then, let us consider, on the same lines of \cite{Concha:2015tla}, the following decomposition of the superalgebra $\mathfrak{g}=\mathfrak{osp}(4|1)$ in three subspaces $V_p$, $p=0,1,2$:
\begin{equation}
\mathfrak{g}= \mathfrak{osp}(4|1) = \mathfrak{so}(3,1) \oplus \frac{\mathfrak{osp}(4|1)}{\mathfrak{sp}(4)}  \oplus \frac{\mathfrak{sp}(4)}{\mathfrak{so}(3,1)} = V_0  \oplus  V_1  \oplus V_2  ,
\end{equation}
where $V_0 = \lbrace \tilde{J}_{ab} \rbrace$, $V_1 = \lbrace \tilde{Q}_\alpha \rbrace$, and $V_2 = \lbrace \tilde{P}_a \rbrace $. Consequently, we can write the subspace structure
\begin{equation}\label{structure}
\begin{split}
[V_0, V_0] & \subset V_0 , \quad [V_1, V_1] \subset V_0 \oplus V_2 {,} \\
[V_0, V_1] & \subset V_1 , \quad [V_1, V_2] \subset V_1 {,} \\
[V_0, V_2] & \subset V_2 , \quad [V_2, V_2] \subset V_0 .
\end{split}
\end{equation}
Now, let us consider the abelian semigroup $\hat{S}= \lbrace \lambda_0, \lambda_1, \lambda_2 , \lambda_3 \rbrace$ described by the multiplication table given in Table \ref{mult}.
\begin{table}[ht]
\centering
\begin{tabular}[t]{c|cccc}
 & $\lambda_0$ & $\lambda_1$ & $\lambda_2$ & $\lambda_3$ \\ 
\hline
$\lambda_0$ & $\lambda_0$ & $\lambda_1$ & $\lambda_2$ & $\lambda_3$ \\ 
$\lambda_1$ & $\lambda_1$ & $\lambda_2$ & $\lambda_3$ & $\lambda_2$ \\ 
$\lambda_2$ & $\lambda_2$ & $\lambda_3$ & $\lambda_2$ & $\lambda_3$ \\ 
$\lambda_3$ & $\lambda_3$ & $\lambda_2$ & $\lambda_3$ & $\lambda_2$ \\ 
\end{tabular} 
\caption{Multiplication table of the semigroup $\hat{S}$.}
\label{mult}
\end{table}

Then, consider the subset decomposition $\hat{S} = \hat{S}_0 \cup \hat{S}_1 \cup \hat{S}_2 $ with
\begin{equation}\label{decomposition}
\hat{S}_0  = \lbrace \lambda_0 , \lambda_2 \rbrace , \quad \hat{S}_1  = \lbrace \lambda_1 , \lambda_3 \rbrace , \quad \hat{S}_2  = \lbrace \lambda_2 \rbrace .
\end{equation}
The decomposition given in \eqref{decomposition} is resonant (see \cite{Izaurieta:2006zz} for details), since it satisfies 
\begin{equation}
\begin{split}
\hat{S}_0 \cdot \hat{S}_0 & \subset \hat{S}_0 , \quad \hat{S}_1 \cdot \hat{S}_1 \subset \hat{S}_0 \cap S_2 {,} \\
\hat{S}_0 \cdot \hat{S}_1 & \subset \hat{S}_1 , \quad \hat{S}_1 \cdot \hat{S}_2 \subset \hat{S}_1 {,} \\
\hat{S}_0 \cdot \hat{S}_2 & \subset \hat{S}_2 , \quad \hat{S}_2 \cdot \hat{S}_2 \subset \hat{S}_0 ,
\end{split}
\end{equation}
which has the same form of \eqref{structure}.
Then, according to Theorem IV.2 of \cite{Izaurieta:2006zz}, the subalgebra
\begin{equation}
\mathfrak{g}_R = W_0 \oplus W_1 \oplus W_2 ,
\end{equation}
where
\begin{equation}
\begin{split}
& W_0 = (\hat{S}_0 \times V_0) = \lbrace \lambda_0 , \lambda_2 \rbrace \times \lbrace \tilde{J}_{ab} \rbrace = \lbrace \lambda_0 \tilde{J}_{ab} , \lambda_2 \tilde{J}_{ab} \rbrace , \\
& W_1 = (\hat{S}_1 \times V_1) = \lbrace \lambda_1 , \lambda_3 \rbrace \times \lbrace \tilde{Q}_\alpha \rbrace = \lbrace \lambda_1 \tilde{Q}_\alpha , \lambda_3 \tilde{Q}_\alpha \rbrace , \\
& W_2 = (\hat{S}_2 \times V_2) = \lbrace \lambda_2 \rbrace \times \lbrace \tilde{P}_a  \rbrace = \lbrace \lambda_2 \tilde{P}_{a} \rbrace ,
\end{split}
\end{equation}
is a resonant subalgebra of $\hat{S} \times \mathfrak{g}$.
We can then perform the following identification:
\begin{equation}
J_{ab}  = \lambda_0 \tilde{J}_{ab} , \quad Z_{ab}  = \lambda_2 \tilde{J}_{ab} , \quad P_a  = \lambda_2 \tilde{P}_a , \quad Q_\alpha  = \lambda_1 \tilde{Q}_\alpha , \quad \Sigma_\alpha  = \lambda_3 \tilde{Q}_\alpha ,
\end{equation}
being $\lbrace J_{ab}, P_a , Z_{ab}, Q_\alpha , \Sigma_\alpha \rbrace$ the set of generators of the new superalgebra obtained after a resonant $\hat{S}$-expansion of $\mathfrak{osp}(4|1)$. This superalgebra actually corresponds to a new supersymmetrization of the $AdS$-Lorentz algebra $AdS-\mathcal{L}_4$ of \cite{Salgado:2014qqa}. Let us call it $\tilde{s}AdS-\mathcal{L}_4$.
Its (anti)commutation relations can be obtained by using the multiplication rules of the semigroup $\hat{S}$ (see Table \ref{mult}) together with the commutation relations \eqref{osp} of $\mathfrak{osp}(4|1)$; they read {as follows}:
\begin{equation}\label{AdSsL}
\begin{split}
\left[ J_{ab},J_{cd}\right] & =\eta _{bc}J_{ad}-\eta _{ac}J_{bd}-\eta
_{bd}J_{ac}+\eta _{ad}J_{bc}, \\
\left[ J_{ab},P_{c}\right] & =\eta _{bc}P_{a}-\eta _{ac}P_{b}, \\
\left[ J_{ab},Z_{cd}\right] & =\eta _{bc}Z_{ad}-\eta _{ac}Z_{bd}-\eta
_{bd}Z_{ac}+\eta _{ad}Z_{bc}, \\
\left[ Z_{ab},Z_{cd}\right] & = \eta _{bc}Z_{ad}-\eta _{ac}Z_{bd}-\eta
_{bd}Z_{ac}+\eta _{ad}Z_{bc},  \\
\left[ Z_{ab},P_{c}\right] & = \eta _{bc}P_{a}-\eta _{ac}P_{b}, \\
\left[ P_a , P_b \right]  & = Z_{ab}, \\
\left[ J_{ab},Q_{\alpha }\right] & = -\frac{1}{2}\left( \gamma _{ab}Q \right)_{\alpha }, \quad 
\left[ J_{ab},\Sigma_{\alpha }\right]  = -\frac{1}{2}\left( \gamma _{ab}\Sigma \right)_{\alpha }, \\
\left[ Z_{ab},Q_{\alpha }\right] & = -\frac{1}{2}\left( \gamma _{ab}\Sigma \right)_{\alpha }, \quad 
\left[ Z_{ab},\Sigma_{\alpha }\right]  = -\frac{1}{2}\left( \gamma _{ab}\Sigma \right)_{\alpha }, \\
\left[ P_{a},Q_{\alpha }\right] & = -\frac{1}{2} \left( \gamma _{a}\Sigma\right) _{\alpha }, \quad 
\left[ P_{a},\Sigma_{\alpha }\right]  = -\frac{1}{2} \left( \gamma _{a}\Sigma\right) _{\alpha }, \\
\left\{ Q_{\alpha },Q_{\beta }\right\} & = - \frac{1}{2} \left[ \left(\gamma^{ab} C \right)_{\alpha \beta} Z_{ab} - 2 \left( \gamma ^{a}C\right) _{\alpha \beta }P_{a} \right] , \\
\left\{ Q_{\alpha },\Sigma_{\beta }\right\} & = - \frac{1}{2} \left[ \left(\gamma^{ab} C \right)_{\alpha \beta} Z_{ab} - 2 \left( \gamma ^{a}C\right) _{\alpha \beta }P_{a} \right] , \\
\left\{ \Sigma_{\alpha },\Sigma_{\beta }\right\} & = - \frac{1}{2} \left[ \left(\gamma^{ab} C \right)_{\alpha \beta} Z_{ab} - 2 \left( \gamma ^{a}C\right) _{\alpha \beta }P_{a} \right] .
\end{split}
\end{equation}
Observe that a new Majorana spinor charge $\Sigma_\alpha$ has been introduced as
a direct consequence of the $\hat{S}$-expansion procedure.

The new $AdS$-Lorentz superalgebra \eqref{AdSsL} contains the so-called $AdS-\mathcal{L}_4$ algebra generated by $\lbrace J_{ab}, P_a, Z_{ab} \rbrace$ as a bosonic subalgebra. The $AdS-\mathcal{L}_4$ algebra and its generalizations have been largely studied and analyzed in \cite{Salgado:2014qqa}. In particular, it was proven that $AdS-\mathcal{L}_4$ allows to include a generalized cosmological term in a Born-Infeld gravity action.
Furthermore, performing the rescaling
\begin{equation}
J_{ab} \rightarrow J_{ab}, \quad Z_{ab} \rightarrow \mu^2 Z_{ab}, \quad P_a \rightarrow \mu P_a 
\end{equation}
and taking the limit $\mu \rightarrow \infty$ (In\"{o}n\"{u}-Wigner contraction) in $AdS-\mathcal{L}_4$, one obtains the minimal Maxwell algebra \eqref{maxwell}.

On the other hand, $AdS-\mathcal{L}_4$ is also a bosonic subalgebra of the $AdS$-Lorentz superalgebra of \cite{Concha:2015tla}. In other words, the new $AdS$-Lorentz superalgebra \eqref{AdSsL} and the $AdS$-Lorentz superalgebra of \cite{Concha:2015tla} share the same bosonic subalgebra $AdS-\mathcal{L}_4$. 

Nevertheless, as we have already mentioned in the Introduction, the $AdS$-Lorentz superalgebra of \cite{Concha:2015tla} has just one fermionic generator.
Differently, $\tilde{s}AdS-\mathcal{L}_4$, given by \eqref{AdSsL}, possesses two fermionic charges $Q_\alpha$ and $\Sigma_\alpha$. Then, since it also contains $AdS-\mathcal{L}_4$ as a bosonic subalgebra, in this sense \eqref{AdSsL} {it} could also be viewed as the minimal supersymmetrization of a minimal Maxwell-like algebra {(that is, actually, a deformation of the Maxwell algebra $\mathcal{M}$, see also \cite{Gomis:2009dm})} in which the bosonic generator $Z_{ab}$ is non-abelian ($\left[ Z_{ab},Z_{cd}\right]  = \eta _{bc}Z_{ad}-\eta _{ac}Z_{bd}-\eta _{bd}Z_{ac}+\eta _{ad}Z_{bc}$) and where $\left[ Z_{ab},P_{c}\right] = \eta _{bc}P_{a}-\eta _{ac}P_{b}$, even if the generators $Z_{ab}$, in this case, do not behave as Lorentz generators when considering the supersymmetric extension and the corresponding commutation relations with the fermionic charges. In this context, we observe that the behavior of the generators $Z_{ab}$ in $\tilde{s}AdS-\mathcal{L}_4$ is also different from the behavior of the $Z_{ab}$'s in (a contraction of) the generalized minimal $AdS$-Lorentz superalgebra of \cite{Concha:2015tla}.

\section{Generalized supersymmetric cosmological term in $D=4$ from $\tilde{s}AdS-\mathcal{L}_4$}\label{S2}

In order to construct an action based on $\tilde{s}AdS-\mathcal{L}_4$ we start, on the same lines of \cite{Concha:2014tca, Concha:2015tla}, from the following $1$-form connection:
\begin{equation}\label{connectionA}
A= A^A T_A = \frac{1}{2} \omega^{ab} J_{ab} + \frac{1}{\ell} V^a P_a + \frac{1}{\ell} k^{ab} Z_{ab} + \frac{1}{\sqrt{\ell}} \psi^\alpha Q_\alpha + \frac{1}{\sqrt{\ell}} \xi^\alpha \Sigma_\alpha ,
\end{equation}
where the $1$-form gauge fields are given by
\begin{equation}
\omega^{ab}  = \lambda_0 \tilde{\omega}^{ab} , \quad V^a  = \lambda_2 \tilde{V}^a , \quad k^{ab}  = \lambda_2 \tilde{\omega}^{ab} , \quad \psi^\alpha  = \lambda_1 \tilde{\psi}^\alpha , \quad \xi^\alpha  = \lambda_3 \tilde{\psi}^\alpha ,
\end{equation}
in terms of the components of the $\mathfrak{osp}(4|1)$ connection
\begin{equation}\label{connosp}
\tilde{A}= \frac{1}{2} \tilde{\omega}^{ab} \tilde{J}_{ab} + \frac{1}{\ell} \tilde{V}^a \tilde{P}_a  + \frac{1}{\sqrt{\ell}} \tilde{\psi}^\alpha \tilde{Q}_\alpha .
\end{equation}

Note that, in order to properly interpret the gauge fields, it is necessary to introduce a length scale $\ell$.
The $1$-forms $\omega^{ab}, V^a, k^{ab}, \psi^\alpha, \xi^\alpha$ are the spin connection, the vielbein, a bosonic $1$-form field, the gravitino field, and an extra spinor $1$-form field, respectively ($\psi^\alpha$ and $\xi^\alpha$ are both Majorana spinors).

The $2$-form curvature $F= dA + A \wedge A$ associated with the connection \eqref{connectionA} reads
\begin{equation}\label{2F}
F = F^A T_A = \frac{1}{2} R^{ab} J_{ab} + \frac{1}{\ell} R^a P_a + \frac{1}{2}F^{ab} k_{ab} + \frac{1}{\sqrt{\ell}} \rho^\alpha Q_\alpha + \frac{1}{\sqrt{\ell}} \Xi^\alpha \Sigma_\alpha ,
\end{equation}
where, in particular, we have\footnote{Here as well as in the sequel, for simplicity, we omit the spinor index $\alpha$, and the wedge product $\wedge$ between differential forms is understood.}
\begin{subequations}
\begin{align}
R^{ab} & = d \omega^{ab} + \omega^a_{\;c} \omega^{cb} , \label{rab} \\
R^a & = d V^a + \omega^a_{\;b} V^b + k^a_{\; b} V^b - \frac{1}{2} \bar{\psi} \gamma^a \psi - \bar{\psi} \gamma^a \xi - \frac{1}{2} \bar{\xi} \gamma^a \xi \nonumber \\
& = D V^a + k^a_{\; b} V^b - \frac{1}{2} \bar{\psi} \gamma^a \psi - \bar{\psi} \gamma^a \xi - \frac{1}{2} \bar{\xi} \gamma^a \xi, \label{ra} \\
F^{ab} & = d k^{ab} + 2 \omega^a_{\;c} k^{cb} + k^a_{\;c} k^{cb} + \frac{1}{\ell^2} V^a V^b + \frac{1}{2 \ell} \bar{\psi} \gamma^{ab} \psi + \frac{1}{\ell} \bar{\psi} \gamma^{ab} \xi + \frac{1}{2\ell} \bar{\xi} \gamma^{ab} \xi \nonumber \\
& = D k^{ab} + k^a_{\;c} k^{cb} + \frac{1}{\ell^2} V^a V^b + \frac{1}{2 \ell} \bar{\psi} \gamma^{ab} \psi + \frac{1}{\ell} \bar{\psi} \gamma^{ab} \xi + \frac{1}{2\ell} \bar{\xi} \gamma^{ab} \xi {,} \label{fab} \\
\rho & = d \psi + \frac{1}{4} \omega^{ab} \gamma_{ab} \psi = D \psi , \label{rho} \\
\Xi & = d \xi + \frac{1}{4} \omega^{ab} \gamma_{ab} \xi + \frac{1}{2 \ell} V^a \gamma_a \psi + \frac{1}{2 \ell} V^a \gamma_a \xi + \frac{1}{4} k^{ab} \gamma_{ab} \psi + \frac{1}{4} k^{ab} \gamma_{ab} \xi \nonumber \\
& = D \xi  + \frac{1}{2 \ell} V^a \gamma_a \psi + \frac{1}{2 \ell} V^a \gamma_a \xi + \frac{1}{4} k^{ab} \gamma_{ab} \psi + \frac{1}{4} k^{ab} \gamma_{ab} \xi , \label{xi} 
\end{align}
\end{subequations}
being $D= d + \omega$ the Lorentz covariant derivative. Setting $F=0$ one retrieves the Maurer-Cartan equations for the superalgebra $\tilde{s}AdS-\mathcal{L}_4$.

Considering the Bianchi identity $\nabla F =0$ ($\nabla = d + [A, \cdot]$), we obtain: 
\begin{subequations}
\begin{align}
D R^{ab} & = 0 , \label{brab} \\
D R^a & = R^a_{\;b}V^b + F^a_{\;b}V^b - k^a_{\;b} R^b + \bar{\psi} \gamma^a \rho + \bar{\psi} \gamma^a \Xi + \bar{\xi} \gamma^a \rho + \bar{\xi} \gamma^a \Xi  , \label{bra} \\
D F^{ab} & = 2 R^a_{\;c} k^{cb} + 2 F^a_{\;c} k^{cb} + \frac{2}{\ell^2} R^a V^b - \frac{1}{\ell} \bar{\psi} \gamma^{ab} \rho - \frac{1}{\ell} \bar{\psi} \gamma^{ab} \Xi - \frac{1}{\ell} {\bar{\xi}} \gamma^{ab} \rho - \frac{1}{\ell} \bar{\xi} \gamma^{ab} \Xi , \label{bfab} \\
D \rho & = \frac{1}{4} R^{ab} \gamma_{ab} \psi , \label{brho} \\
D \Xi & = \frac{1}{4} R^{ab} \gamma_{ab} \xi + \frac{1}{2 \ell} R^a \gamma_a \psi - \frac{1}{2 \ell} V^a \gamma_a \rho + \frac{1}{2 \ell} R^a \gamma_a \xi - \frac{1}{2 \ell} V^a \gamma_a \Xi \nonumber \\
& + \frac{1}{4} F^{ab} \gamma_{ab} \psi - \frac{1}{4} k^{ab} \gamma_{ab} \rho + \frac{1}{4} F^{ab} \gamma_{ab} \xi - \frac{1}{4} k^{ab} \gamma_{ab} \Xi . \label{bxi} 
\end{align}
\end{subequations}

On the other hand, for a clearer understanding of the procedure we are going to follow, let us recall that the $2$-form curvature $\tilde{F}= d\tilde{A} + \tilde{A} \wedge \tilde{A}$ associated with the $\mathfrak{osp}(4|1)$ connection \eqref{connosp} is
\begin{equation}\label{twoformcurvosp}
\tilde{F} = \frac{1}{2} \tilde{R}^{ab} \tilde{J}_{ab} + \frac{1}{\ell} \tilde{R}^a \tilde{P}_a +  \frac{1}{\sqrt{\ell}} \tilde{\rho}^\alpha \tilde{Q}_\alpha ,
\end{equation}
where, as it is well known, we have
\begin{subequations}
\begin{align}
\tilde{R}^{ab} & = d \tilde{\omega}^{ab} + \tilde{\omega}^a_{\;c} \tilde{\omega}^{cb} + \frac{1}{\ell^2} \tilde{V}^a \tilde{V}^b + \frac{1}{2 \ell} \bar{\tilde{\psi}} \gamma^{ab} \tilde{\psi} , \label{rabosp} \\
\tilde{R}^a & = d \tilde{V}^a + \tilde{\omega}^a_{\;b} \tilde{V}^b - \frac{1}{2} \bar{\tilde{\psi}} \gamma^a \tilde{\psi} = \tilde{D} \tilde{V}^a - \frac{1}{2} \bar{\tilde{\psi}} \gamma^a \tilde{\psi},  \label{raosp} \\
\tilde{\rho} & = d \tilde{\psi} + \frac{1}{4} \tilde{\omega}^{ab} \gamma_{ab} \tilde{\psi}  + \frac{1}{2 \ell} \tilde{V}^a \gamma_a \tilde{\psi} = \tilde{D} \tilde{\psi} + \frac{1}{2 \ell} \tilde{V}^a \gamma_a \tilde{\psi} , \label{rhoosp} 
\end{align}
\end{subequations}
being $\tilde{D} = d + \tilde{\omega}$. Now, as recalled in \cite{Concha:2014tca, Concha:2015tla}, the general form of the MacDowell-Mansouri action \cite{MacDowell:1977jt} constructed with the $\mathfrak{osp}(4|1)$ $2$-form curvature $\tilde{F}$ is
\begin{equation}\label{MMosp}
S = 2 \int \langle \tilde{F} \wedge \tilde{F} \rangle = 2 \int \tilde{F}^A \wedge \tilde{F}^B \langle \tilde{T}_A \tilde{T}_B \rangle , 
\end{equation}
with the following choice of the invariant tensor:
\begin{equation}\label{invtenosp}
\langle \tilde{T}_A \tilde{T}_B \rangle = \left\{
\begin{aligned}
\langle \tilde{J}_{ab} \tilde{J}_{cd} \rangle = & \epsilon_{abcd} , \\
\langle \tilde{Q}_\alpha \tilde{Q}_\beta \rangle = & 2 \left( \gamma_5 \right)_{\alpha \beta} .
\end{aligned}
\right.
\end{equation}

Observe that if one chooses the whole $\langle \tilde{T}_A \tilde{T}_B \rangle$ as an invariant tensor (which satisfies
the Bianchi identities) for the $OSp(4|1)$ supergroup, then the action \eqref{MMosp} is a
topological invariant and gives no equations of motion. Nevertheless, with the choice \eqref{invtenosp} of the invariant tensor (which breaks the $OSp(4|1)$ supergroup to its Lorentz subgroup), \eqref{MMosp} becomes a dynamical action that corresponds to the MacDowell-Mansouri action for the $\mathfrak{osp}(4|1)$ superalgebra \cite{MacDowell:1977jt, Castellani:2013iq}.
Writing the explicit form of the action \eqref{MMosp} with the choice \eqref{invtenosp} and omitting the boundary terms, the result is the $\mathcal{N}=1$ supergravity action in four dimensions, given by the Einstein-Hilbert and Rarita-Schwinger terms plus the usual supersymmetric cosmological terms; the aforementioned action is not invariant under the $\mathfrak{osp}(4|1)$ gauge transformations. However, the invariance of the action under supersymmetry transformation can be obtained by modifying the supersymmetry transformation of the spin connection $\tilde{\omega}^{ab}$ \cite{vanNieuwenhuizen:2004rh}.

Now, in order to construct a MacDowell-Mansouri like action for $\tilde{s}AdS-\mathcal{L}_4$, we consider the $\hat{S}$-expansion of $\langle \tilde{T}_A \tilde{T}_B \rangle$ and the $2$-form curvature $F$ in \eqref{2F}.
In particular, the action for $\tilde{s}AdS-\mathcal{L}_4$ can be written as
\begin{equation}\label{AdSsLaction}
S =  2 \int F^A \wedge F^B \langle T_A T_B \rangle ,
\end{equation}
where $\langle T_A T_B \rangle$ can be obtained from the components of the invariant tensor written in \eqref{invtenosp}, using Theorem VII.1 of \cite{Izaurieta:2006zz}. One can then show that the non-vanishing components of $\langle T_A T_B \rangle$ are
\begin{equation}\label{invtensAdSsL}
\begin{split}
& \langle J_{ab} J_{cd} \rangle = C_0 \langle \tilde{J}_{ab} \tilde{J}_{cd} \rangle = C_0 \epsilon_{abcd} , \\
& \langle J_{ab} Z_{cd} \rangle = C_2 \langle \tilde{J}_{ab} \tilde{J}_{cd} \rangle = C_2 \epsilon_{abcd} , \\
& \langle Z_{ab} Z_{cd} \rangle = C_2 \langle \tilde{J}_{ab} \tilde{J}_{cd} \rangle = C_2 \epsilon_{abcd}, \\
& \langle Q_\alpha Q_\beta \rangle = C_2 \langle \tilde{Q}_\alpha \tilde{Q}_\beta \rangle = 2 C_2 \left( \gamma_5 \right)_{\alpha \beta} , \\
& \langle Q_\alpha \Sigma_\beta \rangle = C_2 \langle \tilde{Q}_\alpha \tilde{Q}_\beta \rangle = 2 C_2 \left( \gamma_5 \right)_{\alpha \beta} , \\
& \langle \Sigma_\alpha \Sigma_\beta \rangle = C_2 \langle \tilde{Q}_\alpha \tilde{Q}_\beta \rangle = 2 C_2 \left( \gamma_5 \right)_{\alpha \beta} ,
\end{split}
\end{equation}
where $C_0$ and $C_2$ are (dimensionless) independent constants. 

Thus, considering \eqref{invtensAdSsL} and the $2$-form curvature \eqref{2F}, we can write an action of the form \eqref{AdSsLaction} as
\begin{equation}\label{actioncompact}
S = \int \left( \frac{C_0}{2} \epsilon_{abcd} R^{ab} R^{cd} + C_2 \epsilon_{abcd} R^{ab} F^{cd} + \frac{C_2}{2} \epsilon_{abcd} F^{ab} F^{cd} + \frac{4 C_2}{\ell} \bar{\rho} \gamma_5 \rho + \frac{8 C_2}{\ell} \bar{\rho} \gamma_5 \Xi + \frac{4 C_2}{\ell} \bar{\Xi} \gamma_5 \Xi \right) .
\end{equation}
Then, using the formulas collected in the Appendix and the Bianchi identities originated from the superalgebra $\tilde{s}AdS-\mathcal{L}_4$, one can show that the MacDowell-Mansouri like action \eqref{actioncompact} for $\tilde{s}AdS-\mathcal{L}_4$ can be written explicitly as
\begin{equation}\label{explicitaction}
\begin{split}
S & = \int \Bigg \lbrace \frac{C_0}{2} \epsilon_{abcd} R^{ab} R^{cd} + C_2 \Bigg (  \epsilon_{abcd}  R^{ab} \mathcal{F}^{cd} + \frac{1}{2} \epsilon_{abcd} \mathcal{F}^{ab} \mathcal{F}^{cd} \Bigg ) \\
& + \frac{C_2}{\ell^2} \Bigg ( {\epsilon_{abcd}} R^{ab} V^c V^d + 4 \bar{\psi} V^a \gamma_a \gamma_5 \rho +  4 \bar{\psi}V^a \gamma_a \gamma_5 \sigma  + 4 \bar{\xi}V^a \gamma_a \gamma_5 \sigma + 4 \bar{\xi}V^a \gamma_a \gamma_5 \rho \Bigg ) \\
& + \frac{C_2}{\ell^2} \epsilon_{abcd} \Bigg ( \mathcal{F}^{ab} V^c V^d + \frac{2}{\ell} \bar{\psi} \gamma^{ab} \xi V^c V^d + \frac{1}{\ell} \bar{\xi} \gamma^{ab} \xi V^c V^d + \frac{1}{\ell} \bar{\psi} \gamma^{ab} \psi V^c V^d + \frac{1}{2 \ell^2} V^a V^b V^c V^d  \Bigg ) \\
& + \frac{C_2}{\ell} d  \Bigg ( 4 \bar{\psi} \gamma_5 \rho + 4 \bar{\psi} \gamma_5 \sigma + 4 \bar{\xi} \gamma_5 \rho + 4 \bar{\xi} \gamma_5 \sigma \Bigg ) \Bigg \rbrace ,
\end{split}
\end{equation}
where we have also isolated the boundary terms and defined
\begin{subequations}
\begin{align}
\mathcal{F}^{ab} & = D k^{ab} + k^a_{\;c}k^{cb} , \\
\sigma & = D \xi  + \frac{1}{4} k^{ab} \gamma_{ab} \psi + \frac{1}{4} k^{ab} \gamma_{ab} \xi .
\end{align}
\end{subequations}
Note that we have separated the action \eqref{explicitaction} in different pieces, in such a way to make their physical meaning manifest: The piece proportional to $C_0$ corresponds to the Gauss-Bonnet term; the second term is an Euler invariant and can be seen as a Gauss-Bonnet like term (it does not contribute to the dynamics of the theory and can be written as a boundary term) that involves the new $\tilde{s}AdS-\mathcal{L}_4$ field $k^{ab}$; the third piece contains the Einstein-Hilbert and Rarita-Schwinger Lagrangian, describing pure supergravity, plus three additional terms involving the new spinor $1$-form field $\xi$ and the bosonic field $k^{ab}$; the fourth term corresponds to a generalized supersymmetric cosmological term which contains, besides the usual supersymmetric cosmological term, also two additional terms involving the $\tilde{s}AdS-\mathcal{L}_4$ spinor field $\xi$ and one additional term involving $k^{ab}$; the last term is a boundary term.

Thus, we have shown that the MacDowell-Mansouri like action constructed applying the properties of the $S$-expansion procedure in the case of a resonant $\hat{S}$-expansion (see Table \ref{mult} for the multiplication table of the semigroup $\hat{S}$) of $\mathfrak{osp}(4|1)$ describes a supergravity model with a generalized supersymmetric cosmological term. 
In other words, we have introduced in alternative way the supersymmetric cosmological term in a supergravity model, building a (deformed) $D = 4$ supergravity action from the new $AdS$-Lorentz superalgebra $\tilde{s}AdS-\mathcal{L}_4$. 
Our result corresponds to a new supersymmetric extension of \cite{Salgado:2014qqa} involving two fermionic generators.

Let us observe that, if we consider $k^{ab} =0$ and $\xi =0$ in the action \eqref{explicitaction}, we obtain the MacDowell-Mansouri action for the supergroup $OSp(4|1)$. On the other hand, setting only $\xi=0$ in \eqref{explicitaction}, we obtain the action for the $AdS$-Lorentz superalgebra found in \cite{Concha:2015tla}. 

Notice that if we omit the boundary terms in \eqref{explicitaction}, we get:
\begin{equation}\label{actionnobdy}
\begin{split}
S & = \int \Bigg \lbrace \frac{C_2}{\ell^2} \Bigg ( {\epsilon_{abcd}} R^{ab} V^c V^d + 4 \bar{\psi} V^a \gamma_a \gamma_5 \rho +  4 \bar{\psi}V^a \gamma_a \gamma_5 \sigma  + 4 \bar{\xi}V^a \gamma_a \gamma_5 \sigma + 4 \bar{\xi}V^a \gamma_a \gamma_5 \rho \Bigg ) \\
& + \frac{C_2}{\ell^2} \epsilon_{abcd} \Bigg ( \mathcal{F}^{ab} V^c V^d + \frac{2}{\ell} \bar{\psi} \gamma^{ab} \xi V^c V^d + \frac{1}{\ell} \bar{\xi} \gamma^{ab} \xi V^c V^d + \frac{1}{\ell} \bar{\psi} \gamma^{ab} \psi V^c V^d + \frac{1}{2 \ell^2} V^a V^b V^c V^d  \Bigg )  \Bigg \rbrace .
\end{split}
\end{equation}
Then, if we consider $k^{ab} =0$ and $\xi =0$ in the action \eqref{actionnobdy}, we are left with the Einstein-Hilbert and Rarita-Schwinger Lagrangian plus the usual supersymmetric cosmological term.

Let us now compute the variation of the Lagrangian with respect to the different $\tilde{s}AdS-\mathcal{L}_4$ $1$-form fields in order to obtain the field equations. One can prove that, computing the variation of the Lagrangian with respect to $\omega^{ab}$ and imposing $\delta_\omega \mathcal{L} =0$, we get the following field equation (modulo boundary terms) for the $\tilde{s}AdS-\mathcal{L}_4$ supertorsion:
\begin{equation}\label{eomom}
\epsilon_{abcd} R^c V^d =0 .
\end{equation}
From the variation of the Lagrangian with respect to $k^{ab}$, we obtain the same equation.
On the other hand, computing the variation of the Lagrangian with respect to $V^a$ and imposing $\delta_V \mathcal{L}=0$, we get 
\begin{equation}\label{eomV}
2 \epsilon_{abcd} \left(R^{ab} + F^{ab} \right)V^c + 4 \left( \bar{\psi} + \bar{\xi} \right) \gamma_d \gamma_5 \rho + 4 \bar{\xi} \gamma_d \gamma_5 \left( \rho + \Xi \right) =0 ,
\end{equation}
and from the variation of the Lagrangian with respect to $\psi$, imposing $\delta_\psi \mathcal{L} = 0$, we obtain (modulo boundary terms):
\begin{equation}\label{eompsi}
8 V^a \gamma_a \gamma_5 \left( \rho + \Xi \right) +  4 \gamma_a \gamma_5 \left( \psi + \xi \right) R^a = 0 .
\end{equation}
The variation of the Lagrangian with respect to the spinor $1$-form field $\xi$ leads to the same equation.
Note that the field equations \eqref{eomom}, \eqref{eomV}, and \eqref{eompsi} are similar to those of $\mathfrak{osp}(4|1)$ supergravity, the only differences being related to the presence of the new fields $k^{ab}$ and $\xi$.

Interestingly, we observe that one can define a new bosonic field
\begin{equation}\label{omnewfield}
\hat{\omega}^{ab} \equiv \omega^{ab} + k^{ab},
\end{equation}
{which can be interpreted as an extension of the Riemannian connection $\omega^{ab}_{\mu}dx^\mu$ to a non-Riemannian one with torsion,}
together with the covariant derivative
\begin{equation}
\hat{D} \equiv d + \hat{\omega} ,
\end{equation}
and a new spinor $1$-form field
\begin{equation}\label{psinewfield}
\hat{\psi} \equiv \psi + \xi 
\end{equation}
{implying a redefinition of the gravitino $1$-form field.}
Then, exploiting the new definitions, the equations of motion \eqref{eomom}, \eqref{eomV}, and \eqref{eompsi} become, respectively:
\begin{subequations}
\begin{align}
& \epsilon_{abcd} \hat{R}^c V^d =0 , \label{neweomVa} \\
& 2 \epsilon_{abcd} \hat{R}^{ab} V^c + 4 \bar{\hat{\psi}} \gamma_d \gamma_5 \hat{\rho} =0 , \\
& 8 V^a \gamma_a \gamma_5 \hat{\rho} +  4 \gamma_a \gamma_5 \hat{\psi} \hat{R}^a =0, 
\end{align}
\end{subequations}
where we have also defined
\begin{subequations}
\begin{align}
\hat{R}^{ab} & \equiv d \hat{\omega}^{ab} + \hat{\omega}^a_{\;c} \hat{\omega}^{cb} + \frac{1}{2 \ell} \bar{\hat{\psi}} \gamma^{ab} \hat{\psi} , \label{rabnewfield} \\
\hat{R}^a & \equiv \hat{D} V^a - \frac{1}{2} \bar{\hat{\psi}} \gamma_a \hat{\psi} , \\
\hat{\rho} & \equiv \hat{D} \hat{\psi} + \frac{1}{2 \ell} V^a \gamma_a \hat{\psi} . \label{rhonewfield}
\end{align}
\end{subequations}
These new curvatures have the same form of the $\mathfrak{osp}(4|1)$ ones, and the fields $\lbrace \hat{\omega}^{ab}, V^a, \hat{\psi} \rbrace$ fulfill equations of motion that have the same form of those of the $\mathfrak{osp}(4|1)$ supergravity theory. 
In fact, let us also observe that, at the price of introducing the $1$-form fields $k^{ab}$ and $\xi$ (and the corresponding Maurer-Cartan equations), the $\mathfrak{osp}(4|1)$ superalgebra can be mapped into $\tilde{s}AdS-\mathcal{L}_4$, whereby the spin connection and the gravitino are respectively identified with the Lorentz connection and gravitino of a $D=4$ Minkowski spacetime with vanishing Lorentz curvature and vanishing gravitino super field-strength, albeit with a modification of the supertorsion, the latter being non-vanishing in both cases. This point will be further analyzed in a future work.

{Then, exploiting the definitions \eqref{omnewfield}, \eqref{psinewfield}, \eqref{rabnewfield}, and \eqref{rhonewfield}, the action \eqref{actionnobdy} can be rewritten as follows:}
\begin{equation}\label{newformoftheaction}
{S = \int  \frac{C_2}{\ell^2} \Bigg ( \epsilon_{abcd} \hat{R}^{ab} V^c V^d + 4 \bar{\hat{\psi}} V^a \gamma_a \gamma_5 \hat{\rho} + \frac{1}{\ell} \epsilon_{abcd} \bar{\hat{\psi}} \gamma^{ab} \hat{\psi} V^c V^d + \frac{1}{2 \ell^2}\epsilon_{abcd} V^a V^b V^c V^d  \Bigg )  .}
\end{equation}
{The action \eqref{newformoftheaction} contains the Einstein-Hilbert and Rarita-Schwinger Lagrangian, plus the supersymmetric cosmological term, for the new connection $\hat{\omega}^{ab}$ defined in \eqref{omnewfield} and the new gravitino $1$-form field $\hat{\psi}$ defined in \eqref{psinewfield}, which can be seen as a shifted connection and as a shifted gravitino, respectively. In this sense, our result may also be considered as the specific extension to a non-Riemannian framework determined by the structure of the $\tilde{s}AdS-\mathcal{L}_4$ algebra. Indeed, in this context, the antisymmetry $k^{ab} = - k^{ba}$ implies that we are dealing with an Einstein-Cartan geometry with non-metricity tensor equal to zero, because in \eqref{omnewfield} we have $\hat{\omega}^{(ab)}=0$, while a symmetric part of $\hat{\omega}^{ab}$ would have defined a non-vanishing non-metricity tensor (see, for example, \cite{Hehl:1994ue} for details).}

{On the other hand, let us also observe that an In\"{o}n\"{u}-Wigner contraction of the action \eqref{actionnobdy} leads us to the $D = 4$ pure supergravity action}
\begin{equation}\label{puresugraac}
{S = \int  \frac{C_2}{\ell^2} \Bigg ( \epsilon_{abcd} R^{ab} V^c V^d + 4 \bar{\psi} V^a \gamma_a \gamma_5 \rho  \Bigg ) .}
\end{equation}
{In fact, performing the rescaling}
\begin{equation}\label{resc}
{\omega^{ab} \rightarrow \omega^{ab}, \quad V^a \rightarrow \mu^2 V^a, \quad \psi \rightarrow \mu \psi , \quad k^{ab} \rightarrow \mu^4 k^{ab}, \quad \xi \rightarrow \mu^3 \xi}
\end{equation}
{in \eqref{actionnobdy} and dividing the action by $\mu^4$, the four-dimensional pure supergravity action \eqref{puresugraac} is retrieved by taking the limit $\mu \rightarrow 0$.}

{Finally, allowing also the rescaling}
\begin{equation}
{\ell \rightarrow \mu^2 \ell}
\end{equation}
{on the length scale $\ell$ in the action \eqref{actionnobdy}, together with the rescaling \eqref{resc}, dividing the action by $\mu^4$, and taking the limit $\mu \rightarrow 0$, the $D=4$ supergravity action with supersymmetric cosmological term is retrieved, namely}
\begin{equation}
{S = \int \frac{C_2}{\ell^2} \Bigg ( \epsilon_{abcd} R^{ab} V^c V^d + 4 \bar{\psi} V^a \gamma_a \gamma_5 \rho + \frac{1}{\ell} \epsilon_{abcd} \bar{\psi} \gamma^{ab} \psi V^c V^d + \frac{1}{2 \ell^2} \epsilon_{abcd} V^a V^b V^c V^d  \Bigg ) .}
\end{equation}

\subsection{$\tilde{s}AdS-\mathcal{L}_4$ gauge transformations and supersymmetry invariance}

The gauge transformation of the connection $A$ in \eqref{connectionA} is
\begin{equation}
\delta_\varrho A = D \varrho = d \varrho + [A, \varrho] ,
\end{equation}
where $\rho$ is the $\tilde{s}AdS-\mathcal{L}_4$ gauge parameter
\begin{equation}
\varrho = \frac{1}{2} \varrho^{ab} J_{ab} + \frac{1}{2} \kappa^{ab} Z_{ab} + \frac{1}{\ell} \varrho^a P_a + \frac{1}{\sqrt{\ell}} \epsilon^\alpha Q_\alpha + \frac{1}{\sqrt{\ell}} \varepsilon^\alpha \Sigma_\alpha .
\end{equation}
Then, using
\begin{equation}
\delta \left( A^A T_A \right) = d \varrho + \left[  A^B T_B , \varrho^C T_C \right] ,
\end{equation}
we obtain that the $\tilde{s}AdS-\mathcal{L}_4$ gauge transformations are given by:
\begin{subequations}
\begin{align}
\delta \omega^{ab} & =  D \varrho^{ab} , \\
\delta k^{ab} & = D \kappa^{ab} - 2 \varrho^a_{\;c} k^{cb} + 2 k^a_{\;c} \kappa^{cb} + \frac{2}{\ell^2} V^a \varrho^b - \frac{1}{\ell} \bar{\epsilon} \gamma^{ab} \psi - \frac{1}{\ell} \bar{\epsilon} \gamma^{ab} \xi - \frac{1}{\ell} \bar{\varepsilon} \gamma^{ab} \psi - \frac{1}{\ell} \bar{\varepsilon} \gamma^{ab} \xi  , \\
\delta V^a & = D \varrho^a - \varrho^a_{\;b}V^b + k^a_{\;b} \varrho^b - \kappa^a_{\;b}V^b + \bar{\epsilon} \gamma^a \psi + \bar{\epsilon} \gamma^a \xi + \bar{\varepsilon} \gamma^a \psi + \bar{\varepsilon} \gamma^a \xi  , \\
\delta \psi & = d \epsilon + \frac{1}{4} \omega^{ab} \gamma_{ab} \epsilon - \frac{1}{4} \varrho^{ab} \gamma_{ab} \psi , \\
\delta \xi & = d \varepsilon + \frac{1}{4} \omega^{ab} \gamma_{ab} \varepsilon - \frac{1}{4} \varrho^{ab} \gamma_{ab} \xi - \frac{1}{2 \ell} \varrho^a \gamma_a \psi + \frac{1}{2 \ell} V^a \gamma_a \epsilon - \frac{1}{2 \ell} \varrho^a \gamma_a \xi + \frac{1}{2 \ell} V^a \gamma_a \varepsilon \nonumber \\
& + \frac{1}{4} k^{ab} \gamma_{ab} \epsilon - \frac{1}{4} \kappa^{ab} \gamma_{ab} \psi + \frac{1}{4} k^{ab} \gamma_{ab} \varepsilon - \frac{1}{4} \kappa^{ab} \gamma_{ab} \xi .
\end{align}
\end{subequations} 
Analogously, from the gauge variation of the curvature $F$,
\begin{equation}
\delta_\varrho F = \left[ F, \varrho \right] ,
\end{equation}
we can write the following gauge transformations:
\begin{subequations}
\begin{align}
\delta R^{ab} & = 2 R^a_{\;c} \varrho^{cb} , \\
\delta F^{ab} & = 2 R^a_{\;c} \kappa^{cb} - 2 \varrho^a_{\;c} F^{cb} + 2 F^a_{\;c} \kappa^{cb} + \frac{2}{\ell^2} R^a \varrho^b - \frac{1}{\ell} \bar{\epsilon} \gamma^{ab} \rho - \frac{1}{\ell} \bar{\epsilon} \gamma^{ab} \Xi - \frac{1}{\ell} \bar{\varepsilon} \gamma^{ab} \rho - \frac{1}{\ell} \bar{\varepsilon} \gamma^{ab} \Xi  , \\
\delta R^a & = R^a_{\;b} \varrho^b - \varrho^a_{\;b}R^b + F^a_{\;b} \varrho^b - \kappa^a_{\;b}R^b + \bar{\epsilon} \gamma^a \rho + \bar{\epsilon} \gamma^a \Xi + \bar{\varepsilon} \gamma^a \rho + \bar{\varepsilon} \gamma^a \Xi  , \\
\delta \rho & = \frac{1}{4} R^{ab} \gamma_{ab} \epsilon - \frac{1}{4} \varrho^{ab} \gamma_{ab} \rho , \\
\delta \Xi & = \frac{1}{4} R^{ab} \gamma_{ab} \varepsilon - \frac{1}{4} \varrho^{ab} \gamma_{ab} \Xi - \frac{1}{2 \ell} \varrho^a \gamma_a \rho + \frac{1}{2 \ell} R^a \gamma_a \epsilon - \frac{1}{2 \ell} \varrho^a \gamma_a \Xi + \frac{1}{2 \ell} R^a \gamma_a \varepsilon \nonumber \\
& + \frac{1}{4} F^{ab} \gamma_{ab} \epsilon - \frac{1}{4} \kappa^{ab} \gamma_{ab} \rho + \frac{1}{4} F^{ab} \gamma_{ab} \varepsilon - \frac{1}{4} \kappa^{ab} \gamma_{ab} \Xi .
\end{align}
\end{subequations} 
Although the action \eqref{explicitaction} is built from $\tilde{s}AdS-\mathcal{L}_4$, one can prove that it is not invariant under the $\tilde{s}AdS-\mathcal{L}_4$ gauge transformations. 
Furthermore, if we consider the variation of \eqref{explicitaction} under gauge supersymmetry, we find:
\begin{equation}\label{susytrac}
\delta_{\text{susy}} S = - \frac{4 C_2}{\ell^2} \int R^a  \left( \bar{\rho} + \bar{\Xi} \right)\gamma_a \gamma_5 \epsilon .
\end{equation}
Thus, as in the $\mathfrak{osp}(4|1)$ and super-Poincar\'{e} cases, the action is invariant under gauge supersymmetry imposing the supertorsion constraint
\begin{equation}
R^a =0 .
\end{equation}
However, this causes $\omega^{ab}$ to be expressed in terms of the other fields $V^a , k^{ab}, \psi , \xi$ (second order formalism).

Alternatively, on the same lines of \cite{Concha:2014tca, Concha:2015tla}, one can recover the supersymmetry invariance of the action in the first order formalism by modifying the supersymmetry transformation of the spin connection $\omega^{ab}$.
Indeed, if we consider the variation of the action under an arbitrary $\delta \omega^{ab}$, we find
\begin{equation}
\delta _\omega S = \frac{2 C_2}{\ell^2} \int \epsilon_{abcd} R^a V^b \delta \omega^{cd} ,
\end{equation}
which vanishes for arbitrary $\delta \omega^{ab}$ if $R^a = 0$. Then, it is possible to modify $\delta \omega^{ab}$ with the addition of an extra piece to the gauge transformation in such a way that the variation of the action reads
\begin{equation}
\delta S = - \frac{4 C_2}{\ell^2} \int  R^a \left[ \left( \bar{\rho} + \bar{\Xi} \right) \gamma_a \gamma_5 \epsilon - \frac{1}{2} \epsilon_{abcd} V^b \delta_{\text{extra}}\omega^{cd}  \right] ,
\end{equation}
the supersymmetry invariance being fulfilled when
\begin{equation}
\delta_{\text{extra}}  \omega^{ab}  = 2 \epsilon^{abcd} \left( \bar{\zeta}_{ec} \gamma_d \epsilon+ \bar{\zeta}_{de} \gamma_c \gamma_5 \epsilon - \bar{\zeta}_{cd} \gamma_e \gamma_5 \epsilon \right) V^e ,
\end{equation}
with $\bar{\rho}+ \bar{\Xi} = \bar{\zeta}_{ab} V^a V^b$.
Thus, we can conclude that the action in the first order formalism is invariant under the supersymmetry transformations\footnote{Let us mention that, actually, one could retrieve supersymmetry invariance also by modifying the whole $\delta \left( \omega^{ab} + k^{ab} \right)$, obtaining that the action is invariant under $\delta \left( \omega^{ab} + k^{ab} \right) =  2 \epsilon^{abcd} \left( \bar{\zeta}_{ec} \gamma_d \epsilon+ \bar{\zeta}_{de} \gamma_c \gamma_5 \epsilon - \bar{\zeta}_{cd} \gamma_e \gamma_5 \epsilon \right) V^e - \frac{1}{\ell} \bar{\epsilon} \gamma^{ab} \psi - \frac{1}{\ell} \bar{\epsilon} \gamma^{ab} \xi $, together with the other transformations \eqref{trva} and \eqref{trxi}. The transformations we have written in \eqref{trom}, \eqref{trpsi}, and \eqref{trk} correspond to a particular choice for $\delta \omega^{ab}$ and $\delta k^{ab} $ fulfilling $\delta \left( \omega^{ab} + k^{ab} \right) =  2 \epsilon^{abcd} \left( \bar{\zeta}_{ec} \gamma_d \epsilon+ \bar{\zeta}_{de} \gamma_c \gamma_5 \epsilon - \bar{\zeta}_{cd} \gamma_e \gamma_5 \epsilon \right) V^e - \frac{1}{\ell} \bar{\epsilon} \gamma^{ab} \psi - \frac{1}{\ell} \bar{\epsilon} \gamma^{ab} \xi $.}
\begin{subequations}
\begin{align}
\delta \omega^{ab} & =  2 \epsilon^{abcd} \left( \bar{\zeta}_{ec} \gamma_d \epsilon+ \bar{\zeta}_{de} \gamma_c \gamma_5 \epsilon - \bar{\zeta}_{cd} \gamma_e \gamma_5 \epsilon \right) V^e , \label{trom} \\
\delta k^{ab} & = - \frac{1}{\ell} \bar{\epsilon} \gamma^{ab} \psi - \frac{1}{\ell} \bar{\epsilon} \gamma^{ab} \xi  , \label{trk} \\
\delta V^a & = \bar{\epsilon} \gamma^a \psi + \bar{\epsilon} \gamma^a \xi , \label{trva} \\
\delta \psi & = d \epsilon + \frac{1}{4} \omega^{ab} \gamma_{ab} \epsilon  = D \epsilon , \label{trpsi} \\
\delta \xi & = \frac{1}{2 \ell} V^a \gamma_a \epsilon + \frac{1}{4} k^{ab} \gamma_{ab} \epsilon  . \label{trxi}
\end{align}
\end{subequations} 
Observe that supersymmetry is not a gauge symmetry of the action, and the supersymmetry transformations do not close off-shell, while the $\tilde{s}AdS-\mathcal{L}_4$ gauge transformations close off-shell by construction.

Finally, note that there is also another kind of supersymmetry (i.e.{,} a supersymmetry-like symmetry), related to the fermionic generator $\Sigma_\alpha$. The new supersymmetry-like transformations read as follows:
\begin{subequations}
\begin{align}
\delta \omega^{ab} & = 0, \\
\delta k^{ab} & = - \frac{1}{\ell} \bar{\varepsilon} \gamma^{ab} \psi - \frac{1}{\ell} \bar{\varepsilon} \gamma^{ab} \xi  , \\
\delta V^a & = \bar{\varepsilon} \gamma^a \psi + \bar{\varepsilon} \gamma^a \xi  , \\
\delta \psi & = 0 , \\
\delta \xi & = d \varepsilon + \frac{1}{4} \omega^{ab} \gamma_{ab} \varepsilon + \frac{1}{2 \ell} V^a \gamma_a \varepsilon + \frac{1}{4} k^{ab} \gamma_{ab} \varepsilon = D \varepsilon  + \frac{1}{2 \ell} V^a \gamma_a \varepsilon + \frac{1}{4} k^{ab} \gamma_{ab} {\varepsilon} .
\end{align}
\end{subequations}
If we now consider the variation of the action under the new supersymmetry-like transformations, we get:
\begin{equation}
\delta_{\text{susy-like}} S = - \frac{4 C_2}{\ell^2} \int R^a  \left( \bar{\rho} + \bar{\Xi} \right)\gamma_a \gamma_5 \varepsilon ,
\end{equation}
which has the same form of \eqref{susytrac}, the only difference relying in the parameter $\varepsilon$.
Then, one can repeat the procedure described above, obtaining that the action in the first order formalism is invariant under the following new supersymmetry-like transformations:\footnote{Again, one could retrieve the invariance also by modifying the whole $\delta \left( \omega^{ab} + k^{ab} \right)$, obtaining $\delta \left( \omega^{ab} + k^{ab} \right) =  2 \epsilon^{abcd} \left( \bar{\zeta}_{ec} \gamma_d \epsilon+ \bar{\zeta}_{de} \gamma_c \gamma_5 \epsilon - \bar{\zeta}_{cd} \gamma_e \gamma_5 \epsilon \right) V^e - \frac{1}{\ell} \bar{\varepsilon} \gamma^{ab} \psi - \frac{1}{\ell} \bar{\varepsilon} \gamma^{ab} \xi$, together with \eqref{trvanew}, \eqref{trpsinew}, and \eqref{trxinew}.}
\begin{subequations}
\begin{align}
\delta \omega^{ab} & = 2 \epsilon^{abcd} \left( \bar{\zeta}_{ec} \gamma_d \epsilon+ \bar{\zeta}_{de} \gamma_c \gamma_5 \epsilon - \bar{\zeta}_{cd} \gamma_e \gamma_5 \epsilon \right) V^e , \\
\delta k^{ab} & = - \frac{1}{\ell} \bar{\varepsilon} \gamma^{ab} \psi - \frac{1}{\ell} \bar{\varepsilon} \gamma^{ab} \xi  , \\
\delta V^a & = \bar{\varepsilon} \gamma^a \psi + \bar{\varepsilon} \gamma^a \xi  , \label{trvanew} \\
\delta \psi & = 0 , \label{trpsinew} \\
\delta \xi & = d \varepsilon + \frac{1}{4} \omega^{ab} \gamma_{ab} \varepsilon + \frac{1}{2 \ell} V^a \gamma_a \varepsilon + \frac{1}{4} k^{ab} \gamma_{ab} \varepsilon = D \varepsilon + \frac{1}{2 \ell} V^a \gamma_a \varepsilon + \frac{1}{4} k^{ab} \gamma_{ab} {\varepsilon} , \label{trxinew}
\end{align}
\end{subequations}
{being} $\bar{\rho}+ \bar{\Xi} = \bar{\zeta}_{ab} V^a V^b$. Also this supersymmetry-like symmetry is not a gauge symmetry of the action. 
{Further investigation of this symmetry in the context of supergravity theories are currently under analysis (work in progress).}

\section{Discussion}\label{Discussion}

{The Maxwell and $AdS$-Lorentz algebras of all types, together with their supersymmetric extensions, have found interesting applications in (super)gravity, although the new generators (and the consequent introduced modifications) still require a clearer physical interpretation, particularly concerning the presence of extra fermionic generators in the supersymmetric cases (for some progress achieved in this context, see \cite{Ravera:2018vra}).}

In this paper, driven by the fact that from $AdS$-Lorentz type (super)algebras one can introduce the cosmological term in (super)gravity in the presence of an extra bosonic generator $Z_{ab}$ \cite{Salgado:2014qqa, Concha:2015tla}, we have presented a new supersymmetrization of the $AdS$-Lorentz algebra $AdS-\mathcal{L}_4$ of \cite{Salgado:2014qqa}. Compared to previous literature, the novel superalgebra contains two fermionic generators rather than one and, interestingly, it {allows} for non-abelian charges in the underlying Maxwell algebra {(see also \cite{Gomis:2009dm, Bonanos:2010fw} for non-abelian deformations of Maxwell (super)algebras)}.
Specifically, our new $AdS$-Lorentz superalgebra (that we called $\tilde{s}AdS-\mathcal{L}_4$) possesses, by construction, a bosonic subalgebra (that is $AdS-\mathcal{L}_4$) that does not contains any additional generator with respect to the Maxwell algebra $\mathcal{M}$, recalled in \eqref{maxwell}, in such a way that, when contracted, it directly reproduces the Maxwell algebra $\mathcal{M}$.
In this sense, $\tilde{s}AdS-\mathcal{L}_4$ could also be viewed as the minimal supersymmetrization of a minimal Maxwell-like algebra in which the bosonic generator $Z_{ab}$ is non-abelian ($\left[ Z_{ab},Z_{cd}\right]  = \eta _{bc}Z_{ad}-\eta _{ac}Z_{bd}-\eta _{bd}Z_{ac}+\eta _{ad}Z_{bc}$) and where $\left[ Z_{ab},P_{c}\right] = \eta _{bc}P_{a}-\eta _{ac}P_{b}$, even if the generators $Z_{ab}$, in this case, do not behave as Lorentz generators when considering the supersymmetric extension and the corresponding commutation relations with the fermionic charges.

In particular, we have obtained $\tilde{s}AdS-\mathcal{L}_4$ as an $S$-expansion of the superalgebra $\mathfrak{osp}(4|1)$, using the semigroup described by Table \ref{mult}, and, exploiting some peculiar and useful properties of the abelian semigroup expansion method (on the same lines of \cite{Concha:2014tca, Concha:2015tla}), we have shown that it allows to construct in a geometric way a $D=4$ supergravity model involving a generalized supersymmetric cosmological term. The action we have obtained with this procedure corresponds to a MacDowell-Mansouri like action.
Our result is a new supersymmetric extension of \cite{Salgado:2014qqa} involving two fermionic generators.
{Moreover, we have shown that the final action (omitting the boundary terms) can be rewritten as the Einstein-Hilbert and Rarita-Schwinger action, plus the supersymmetric cosmological term, for a new connection $\hat{\omega}^{ab}= \omega^{ab} + k^{ab}$ and gravitino $\hat{\psi}=\psi + \xi$ (that is, a shifted connection and gravitino, respectively). Then, the antisymmetry $k^{ab} = - k^{ba}$ implies that we are dealing with an Einstein-Cartan geometry (determined by the structure of the $\tilde{s}AdS-\mathcal{L}_4$ algebra and involving a redefinition of the gravitino $1$-form field, since we are working in superspace) with vanishing non-metricity.}

Interestingly, in our model the {bosonic $1$-form field $k^{ab}$ (associated to $Z_{ab}$)} and the spinor $1$-form field $\xi$ (associated to
the fermionic charge $\Sigma$) appear not only in the boundary terms but also in the bulk
Lagrangian (analogously to what happened in \cite{Concha:2015tla} and differently from what happened in \cite{deAzcarraga:2014jpa, Concha:2014tca}).
{In particular, the presence of the fields $k^{ab}$ and $\xi$ in the boundary could be useful in the context of the $AdS$/CFT duality (see \cite{Maldacena:1997re, Maldacena:1998, Gubser:1998bc, Witten:1998qj, Aharony:1999ti, DHoker:2002nbb} and references therein). Interestingly, as shown in \cite{Miskovic:2009bm}, the introduction of a topological boundary in a bosonic action in four-dimensions is equivalent to the holographic renormalization procedure (for a review of the holographic renormalization, see, for example, \cite{Skenderis:2002wp}) in the $AdS$/CFT context.
Then, we conjecture that the presence of the fields $k^{ab}$ and $\xi$ in the boundary would allow to regularize the supergravity action in the holographic renormalization context (some work is in progress on this point).}
{On the other hand, at the purely supergravity level, it was shown, adopting the so-called rheonomic (geometric) approach, that the supersymmetry invariance of different supergravity actions in the presence of a non-trivial boundary of spacetime can be recovered by adding appropriate boundary terms, reproducing MacDowell-Mansouri like actions \cite{Andrianopoli:2014aqa, Ipinza:2016con, Banaudi:2018zmh, Concha:2018ywv}. Then, one could investigate the possibility of obtaining the action \eqref{explicitaction} from the rheonomic approach adopted in \cite{Andrianopoli:2014aqa, Ipinza:2016con, Banaudi:2018zmh, Concha:2018ywv} in the presence of a non-trivial boundary of spacetime, where we conjecture that the presence of the fields $k^{ab}$ and $\xi$ in the boundary would allow to recover the supersymmetry invariance of the action. These analyses could also shed some light on the physical interpretation of the new generators and of the induced modifications in the theory appearing when (Maxwell and) $AdS$-Lorentz type algebras are considered.}

{It would also be interesting} to carry on an analysis in $2+1$ dimensions {(in the context of Chern-Simons theories)}, on the same lines of \cite{Concha:2018jxx}, considering the restriction to three dimensions of $\tilde{s}AdS-\mathcal{L}_4$.

On the other hand, a possible future development could consist in analyzing $\mathcal{N}$-extended versions of $\tilde{s}AdS-\mathcal{L}_4$ and constructing $\mathcal{N}$-extended supergravity models (also higher-dimensional and matter-coupled ones) from the aforementioned $\mathcal{N}$-extended superalgebras. In this context, the $S$-expansion procedure could play an important role. In fact, as one can see from the results we have obtained in this paper, the $S$-expansion procedure is not only a useful mathematical method to derive new Lie (super)algebras but also a powerful tool in order to construct, within a geometric formulation, a (super)gravity action for an $S$-expanded (super)algebra. This can give rise to many new extensions and generalizations of the results obtained here and in the literature, as well as to a clearer understanding of the algebraic and physical relations among different theories.

\section{Acknowledgments}

D.M.P. acknowledges DI-VRIEA for financial support through Proyecto Postdoctorado
2018 VRIEAPUCV. The authors wish to thank P. Concha and E. Rodr\'{i}guez for the illuminating discussions.

This is a preprint of the article published in Eur.\ Phys.\ J.\ C {\bf 78} (2018) no.11,  945. The final authenticated version is available online at: https://doi.org/10.1140/epjc/s10052-018-6421-9.

\section{Conventions and useful formulas}\label{appa}

For the Minkowski metric we adopt the convention $\eta_{ab} \equiv (-1,1,1,1)$.
The Dirac gamma matrices in four spacetime dimensions are defined through $\lbrace{ \gamma_a , \gamma_b }\rbrace = - 2 \eta_{ab} $ and obey the following relations:
\begin{equation}\label{gammaform}
\begin{split}
& [\gamma_a , \gamma_b]  = 2 \gamma_{ab} , \; \gamma_5  = - \gamma_0 \gamma_1 \gamma_2 \gamma_3 , \; \gamma^2_5 = -1 , \; \lbrace{ \gamma_5 , \gamma_a }\rbrace =0, \; [\gamma_5 , \gamma_{ab}] =0, \; \gamma_{ab} \gamma_5 = -\frac{1}{2}\epsilon_{abcd} \gamma^{cd}, \\
& \gamma_a \gamma_b  = \gamma_{ab}-\eta_{ab}, \; \gamma^{ab} \gamma_{cd} = \epsilon^{ab}_{\;\;\;cd} \gamma_5 - 4  \delta^{[a}_{\;\;[c} \gamma^{b]}_{\;\;d]}- 2 \delta^{ab}_{cd}, \; \gamma^{ab} \gamma^c = 2 \gamma^{[a}\delta^{b]}_c - \epsilon^{abcd} \gamma_5 \gamma_d , \\
& \gamma^c \gamma^{ab} = - 2 \gamma^{[a}\delta^{b]}_c - \epsilon^{abcd}\gamma_5 \gamma_d , \; \gamma_m \gamma^{ab} \gamma^m  = 0, \; \gamma_{ab} \gamma_m \gamma^{ab} =0, \; \gamma_{ab}\gamma_{cd}\gamma^{ab} = 4 \gamma_{cd}, \; \gamma_m \gamma^a \gamma^m  = -2 \gamma^a . 
\end{split}
\end{equation}
Furthermore, {we have}
\begin{equation}\label{gammasymm}
(C \gamma_a)^T = C \gamma_a , \quad (C \gamma_{ab})^T  = C \gamma_{ab} , \quad  (C \gamma_5)^T = - C \gamma_5 , \quad (C \gamma_5 \gamma_a)^T  = - C \gamma_5 \gamma_a  , 
\end{equation}
where $C$ ($C^T = - C$) is the charge conjugation matrix. 
For a generic Majorana spinor $p$-form $\eta$ ($\bar{\eta} = \eta^T C$) and a generic Majorana spinor $q$-form $\chi$ ($\bar{\chi} = \chi^T C$), the following identities hold:
\begin{equation}\label{spinsymm}
\bar{\eta} \chi = (-1)^{pq}\bar{\chi} \eta , \quad \bar{\eta}  S \chi = - (-1)^{pq} \bar{\chi}  S \eta , \quad \bar{\eta}  A \chi  = (-1)^{pq}\bar{\chi}  A \eta , 
\end{equation}
where $S$ and $A$ are symmetric and a antisymmetric matrices, respectively. 
Finally, in $\mathcal{N}=1$, $D=4$ we can write the following relevant Fierz identities:
\begin{equation}\label{fierzids}
\begin{split}
& \psi \bar{\psi} = \frac{1}{2} \gamma_a \bar{\psi}  \gamma^a \psi - \frac{1}{8} \gamma_{ab} \bar{\psi}  \gamma^{ab} \psi , \\
& \gamma_a \psi  \bar{\psi}  \gamma^a \psi = 0, \\
& \gamma_{ab} \psi  \bar{\psi} \gamma^{ab} \psi = 0, \\
& \gamma_{ab} \psi  \bar{\psi} \gamma^a \psi = \psi  \bar{\psi}  \gamma_b \psi . 
\end{split}
\end{equation}

\end{document}